\documentclass[sigconf]{acmart}

\AtBeginDocument{%
  \providecommand\BibTeX{{%
    \normalfont B\kern-0.5em{\scshape i\kern-0.25em b}\kern-0.8em\TeX}}}

\usepackage{makecell}
\usepackage{threeparttable}
\usepackage{color}
\copyrightyear{2022}
\acmYear{2022}
\setcopyright{acmcopyright}
\acmConference[SIGIR '22]{Proceedings of the 45th
International ACM SIGIR Conference on Research and Development in Information
Retrieval}{July 11--15, 2022}{Madrid, Spain}
\acmBooktitle{Proceedings of the 45th International ACM SIGIR Conference on
Research and Development in Information Retrieval (SIGIR '22), July 11--15, 2022,
Madrid, Spain}
\acmPrice{15.00}
\acmDOI{10.1145/3477495.3532005}
\acmISBN{978-1-4503-8732-3/22/07}

\begin{document}
\fancyhead{}

\title{Investigating Accuracy-Novelty Performance \\ for Graph-based Collaborative Filtering}
\author{Minghao Zhao$^1$, Le Wu$^{2,3, *}$, Yile Liang$^1$, Lei Chen$^{2,3}$, Jian Zhang$^4$, Qilin Deng$^{1,5}$, Kai Wang$^1$} 
\author{Xudong Shen$^1$, Tangjie Lv$^1$, Runze Wu$^{1, *}$}

\affiliation{%
  \institution{$^1$Fuxi AI Lab, NetEase Games, Hangzhou, China}
  \institution{$^2$Hefei University of Technology, Hefei, China}
  \institution{$^3$Institute of Artificial Intelligence, Hefei Comprehensive National Science Center, Hefei, China}
\institution{$^4$Institute of Cyberspace Security, Zhejiang University of Technology, Hangzhou, China}
\country{$^1$\{zhaominghao, liangyile, wangkai02, hzshenxudong, hzlvtangjie, wurunze1\}@corp.netease.com\\ $^2$\{lewu.ustc, chenlei.hfut\}@gmail.com, $^4$jianzh@zjut.edu.cn}, $^5$qlqldeng@163.com
}









\begin{abstract}
Recent  years  have  witnessed  the great  accuracy performance of graph-based Collaborative Filtering~(CF) models for recommender systems. By taking the user-item interaction behavior as a graph, these graph-based CF models borrow the success of Graph Neural Networks (GNN), and iteratively perform neighborhood aggregation to propagate the collaborative signals. While  conventional  CF models are known for facing the challenges of the popularity bias that favors popular items, one may  wonder  “\emph{Whether  the  existing  graph-based  CF models  alleviate or exacerbate popularity bias of recommender systems?}” To answer this question, we first investigate the two-fold performances w.r.t. accuracy and novelty for existing  graph-based CF methods.  The empirical results show that \emph{symmetric neighborhood aggregation adopted by most existing  graph-based  CF models exacerbate the  popularity bias and this phenomenon becomes more serious as the depth of graph propagation increases}. Further, we theoretically  analyze the cause of popularity bias for graph-based CF. Then, we propose a simple yet effective plugin, namely $r$-$AdjNorm$, to achieve an accuracy-novelty trade-off  by controlling the normalization strength in the neighborhood aggregation process. Meanwhile, $r$-$AdjNorm$ can be smoothly applied to the existing graph-based CF backbones without  additional computation. Finally, experimental results on three benchmark datasets show that our proposed method can improve novelty without sacrificing  accuracy under various graph-based CF backbones.

\end{abstract}
                                

\maketitle
\renewcommand{\shortauthors}{Minghao Zhao, et al.}

\newcommand\blfootnote[1]{%
	\begingroup
	\renewcommand\thefootnote{}\footnote{#1}%
	\addtocounter{footnote}{-1}%
	\endgroup
}

\blfootnote{*Corresponding authors.}


\section{Introduction}


Collaborative Filtering (CF) based recommendation  assumes a user is more likely to be interested in items that are enjoyed by other users, and has been widely studied in both academia and industry due to its easy-to-collect data and relatively high recommendation performance~\cite{koren2015advances}. Among all models for CF, Matrix Factorization (MF) learns user and item embeddings by projecting both users and items into a low latent space for user preference prediction~\cite{koren2009matrix,rendle2009bpr,he2017neural}.
Recently, with the huge success of Graph Neural Networks (GNN) for modeling graph-structured data~\cite{kipf2017semi,ying2018graph,tao2019mvan}, researchers argued that user-item behavior data can be naturally formed as a user-item bipartite graph structure. 
Researchers proposed to borrow the ideas of GNN into CF, and various graph-based CF models have been proposed.  These graph-based CF iteratively performs Neighborhood Aggregation (NA) to capture the higher-order collaborative signals by stacking multiple graph convolutional layers~\cite{kipf2017semi}. Therefore, these graph-based CF models can naturally inherit the core idea of CF, and alleviate the data sparsity issue via higher-order graph structure~\cite{berg2017graph,wang2019neural,he2020lightgcn, wu2019neural}. It is now generally accepted that these graph-based CF models achieve State-Of-The-Art (SOTA) recommendation accuracy. 

As recommender systems provide personalized suggestions to increase user satisfaction and platform prosperity,  optimizing the recommendation accuracy is obviously not the single goal. Among all recommendation metrics, the novelty of the recommendation list has also received substantial attention~\cite{zhou2010solving,yin2012challenging, kaminskas2016diversity}. A novel recommended item is one that is previously unpopular and is probably unknown to users.  Therefore, novelty is commonly negatively correlated  with an item's popularity~\cite{castells2015novelty}. In other words,  the less an item is noticed by other users, the more novel  it is. However, it is well known that traditional CF methods  often  unintentionally face the challenge of the novelty recommendation~\cite{yin2012challenging, abdollahpouri2017controlling,canamares2018should}. This is due to the unbalanced long-tail distribution of the observed user-item interaction records, and most CF optimization algorithms would amplify the unbalanced item popularity distribution by focusing on the observed ratings. Therefore, CF algorithms would enlarge popular bias in the data, making the recommendation list more unbalanced by suggesting popular items. Prior works have also shown that popularity bias can be problematic since it impedes the system diversity, 
leading to monotonous recommendations \cite{fleder2009blockbuster,brynjolfsson2006niches}.

The general consensus among researchers of the reason for the popularity bias is due to the imbalanced user-item interaction data, and the amplification effect of CF algorithms. First, the long-tail distribution of item popularity is a common observed phenomenon  from user-item behavior data~\cite{abdollahpouri2017controlling,zhao2021bilateral}. Second, most CF algorithms, especially the popular MF-based algorithms, such as the point-wise based rating loss function~\cite{hu2008collaborative} and the pair-wise implicit feedback based loss function~\cite{rendle2009bpr}, would emphasize on optimizing the observed user interaction records compared to the novel ones. Therefore, instead of recommending novel items that rarely appear in the observed user-item training data, the accuracy-based CF algorithms would amplify the popularity bias and make ``safe'' recommendations that cater to popular items. To improve the novelty performance of recommendation, various novelty enhanced models have been proposed ,  such as re-sampling~\cite{mikolov2013distributed}, re-weighting~\cite{steck2011item}, regularization~\cite{abdollahpouri2017controlling} and re-ranking~\cite{zhu2021popularity}. In general, most of them can be viewed as rebalancing strategies that propel the models to focus more on  optimization of long-tail items. Likewise, researchers have recently started investigating the beyond-accuracy performance of graph-based CF~\cite{sun2020framework, zheng2021dgcn,isufi2021accuracy, wu2021self, wu2021learning}. 
Most of these models focused on diversity metrics, and used heuristics to select neighbors in the NA process to enhance recommendation diversity.
However, they do not focus on novelty and the proposed heuristics need to be carefully designed without any  theoretical guarantee. 
Besides, some causality based popularity debiasing models are also proposed to eliminate the  impact  of item popularity on user interest~\cite{bonner2018causal, zheng2021disentangling,wei2021model}. 


Due to the SOTA recommendation accuracy  of graph-based CF algorithms, in this paper, 
we attempt to investigate  novelty performance under graph-based CF backbones. A preliminary question is : ``\emph{Whether  the  existing graph-based CF models alleviate or exacerbate popularity bias of recommender systems}?''
Some may think that graph-based CF would alleviate popularity bias as the NA process would explore more neighbors to access novel items, while others argue that it would exacerbate popular bias as popular items are more likely to be connected in the aggregation process. To answer this question, we first perform extensive experiments to evaluate the two-fold performances w.r.t accuracy and  novelty of the several existing SOTA graph-based CF methods on three  benchmark datasets. The experimental empirical results~(See Section \ref{sec:pop}) show that the existing graph-based models  exacerbate the popularity bias in the recommender system. Further, we  find that stacking deep graph layers would damage accuracy and novelty simultaneously.

Based on the empirical experimental observations, we  attempt to analyze the cause of popularity bias in graph-based CF model.  We theoretically prove that the normalization coefficient of NA layer plays an important role in controlling the model's performance of item popularity.  Thus, we propose a simple yet effective plugin, i.e., $r$-$AdjNorm$, which is implemented by adjusting the normalization form of NA functions in the exiting graph-based backbones. This plugin makes our research easy to reproduce\footnote{The code and data are available at \url{https://github.com/fuxiAIlab/r-AdjNorm}}. To summarize, the main contributions of this paper are listed as follows:
\begin{itemize}
    \item We study the  accuracy-novelty performance of the  graph-based CF and empirically find that the  existing models with the vanilla GNN layer suffer the popularity bias in recommendation, resulting in poor novelty performance.
    \item We theoretically analyze the cause of popularity bias for graph-based CF, and  propose an  effective method,  $r$-$AdjNorm$, which can be plugged into  existing  graph-based backbones and flexibly control the accuracy-novelty trade-off.
    \item Extensive experiments have  validated the effectiveness and generality of our proposed method in terms of promoting novelty without sacrificing accuracy  compared to a wide range of baselines under various graph-based backbones. 
\end{itemize}

\section{RELATED WORK}


\textbf{Multi-perspective Evaluation of the Recommender Systems}. In the early days, accuracy is commonly used to evaluate the quality of  recommender systems~\cite{wu2022survey}. In addition to accuracy, diversity and novelty have also been extensively studied  since one single metric cannot comprehensively evaluate  the performances of recommendations~\cite{fleder2009blockbuster,castells2015novelty}.  Diversity and novelty are two related but different beyond-accuracy metrics~\cite{vargas2011rank}. The common categories for measuring diversity is intra-user~\cite{zhang2008avoiding} and inter-user level~\cite{adomavicius2011improving}. 
Intra-user level evaluates the dissimilarity of all item pairs in the user's recommendation list, while inter-user level considers the aggregate  diversity of recommendations across all users. 
Novelty reflects the properties  that the recommended items are unknown to users and is usually estimated as negatively correlated with popularity~\cite{castells2011novelty,kaminskas2016diversity} . That is, the more popular the item, the less novel it is to the users. In this paper, we focus on the novelty recommendation.


\textbf{Conventional Novelty Recommendation}. The Bayesian Personalized Ranking (BPR)~\cite{rendle2009bpr}  is extensively applied in MF-based CF by learning pairwise preferences between positive and negative items. However, it still suffers the  popularity bias due to unbalanced data. Thus,  the most widely adopted methods for improving novelty are to design re-balanced strategies, such as re-sampling~\cite{chawla2002smote, mikolov2013distributed} and  re-weighting~\cite{zhou2010solving,steck2011item}. 
For example, 
Lo et al. proposed a personalized pairwise novelty weighting mechanism into the loss function to capture personal preferences for novel items~\cite{lo2019matching}. In addition,  regularization-based methods~\cite{kiswanto2018fairness}, which inject novelty objectives into the loss functions, have also been verified by practical applications. Abdollahpouri et al. introduced a regularization-based framework that encourages items to span different popularity partitions  to improve the long-tail coverage of recommendation lists~\cite{abdollahpouri2017controlling}. Other methods, e.g.,  re-ranking ~\cite{carbonell1998use, abdollahpouri2019managing, zhu2021popularity}, as a post-processing step applied to the basic recommendation lists, the main idea is  to reorder the  lists by taking into  account diversity or novelty metrics.   

\textbf{Popularity Bias and Debiasing}. Another  line of related research  is to eliminate the popularity bias  that popular items are more likely to be recommended to customers than long-tail/niches ones~\cite{celma2008hits,abdollahpouri2019unfairness,borges2020measuring}. Recently, Zhu et al. proposed a new popularity-opportunity bias from the perspective of both  user- and item-side and  theoretically analyzed the influence of item popularity on ranking by MF-based CF~\cite{zhu2021popularity}. They also proposed two metrics, i.e., PRU and PRI, to measure this bias. Zheng et al. leveraged a causal graph  to decompose observed user-item interactions into  user interest and conformity~\cite{zheng2021disentangling}. Through disentangling the representations into these two orthogonal factors, the true interests of users can be learned to debias the impacts of conformity. Likewise, Wei et al. performed multi-task learning to capture different effects of each cause and employed counterfactual reasoning to  eliminate the popularity bias during testing~\cite{wei2021model}. 
Although popularity debiasing and novelty recommendation are correlated in many contexts, they own different research paradigms and  optimization objectives. 
Novelty recommendation usually takes data-driven methods to improve the ability for retrieving novel items, thereby  avoiding recommending  popular items too frequently.  While popularity debiasing methods study how to eliminate the item popularity  for uncovering the true  user preferences from a causal perspective and they are usually evaluated in the debiased dataset to eliminate the influence of popularity  with non IID distributed data~\cite{bonner2018causal, zheng2021disentangling,wei2021model}. Our study is complementary to theirs as we assume the IID distribution of the training and test data, and our focus is to alleviate the popularity bias brought by  graph-based CF models for enhancing  novelty.


\textbf{Multi-objective Tasks for  GNN-based CF.} 
Beyond pursuing higher accuracy, there are some studies on improving the multi-objective performances for the graph-based CF~\cite{sun2020framework, wu2021learning}. 
For example, Sun et al. proposed a framework for improving accuracy and diversity of recommendation by jointly training the model on the observed graph and sampled subgraphs under the Bayesian framework~\cite{sun2020framework}.
Zheng et al. proposed  re-balanced neighbor discovering, category-boosted negative sampling and adversarial learning  to underpin the intra-user diversified recommendation with GNN~\cite{zheng2021dgcn}. Isufi et al. proposed to learn node representations from the nearest  and furthest neighbor graphs jointly to achieve accuracy-diversity trade-off~\cite{isufi2021accuracy}. Most recently, SGL is proposed to leverage self-supervised learning to improve the performance of long-tail items~\cite{wu2021self}. However, there are few studies on  investigation the two-fold performances w.r.t accuracy and novelty of the existing graph-based CF. Our work differs from the above works in that we investigate  the popularity bias caused by deep GNN at a theoretical level, and accordingly propose an efficient method to improve the recommendation novelty based solely on user-item interactions.

\section{PRELIMINARIES}
\subsection{MF-based and Graph-based CF Outlines}
We begin with a brief review of vanilla MF-based CF and then outline the common framework of graph-based CF methods. Let $\mathcal{U}$ and $\mathcal{I}$ denote the set of users and items in the recommender systems, respectively. In this work, we consider the implicit recommendation  since implicit feedbacks are the most common datasets. Let $\mathcal{O^+}=\{y_{ui}|u\in\mathcal{U}, i\in\mathcal{I}\}$ denote the interaction records   observed  between the users and the items, where $y_{ui}=1$ means that the user $u$ has interacted with item $i$. Most existing CF methods treat  user-item interactions ${\mathcal{O^+}}$ as  matrix form, i.e., $ \mathbf{B} \in \mathbb{R}^{|\mathcal{U}|\times|\mathcal{I}|}$, where $|\mathcal{U}|$ and $|\mathcal{I}|$ are the quantity of users and items and each entry $y_{ui}=(\mathbf{B})_{u,i}$  is a binary number, indicating whether the user $u$ has interacted with the item $i$. The user-item interaction matrix $\mathbf{B}$ is usually highly sparse and biased because observed interactions are  scarce compared to the unobserved user-item pairs and popular items account for the majority of interactions.

Both MF-based and graph-based CF aim to learn user and item representations from the interaction data ${\mathcal{O^+}}$. Let $\mathbf{E}\in\mathbb{R}^{(|\mathcal{U}|+|\mathcal{V}|) \times d}$ denote the embedding matrix of users and items, where $\mathbf{E}_{[1:|\mathcal{U}|]}$ is the user sub-matrix and  $\mathbf{E}_{[|\mathcal{U}|+1:|\mathcal{U}|+|\mathcal{V}|]}$ is the item sub-matrix and $d$ is the latent dimension, with $d\ll|\mathcal{U}|, |\mathcal{I}|$. 
The  MF-based CF methods~\cite{koren2009matrix, rendle2009bpr} work by approximately decomposing  the user-item interaction matrix $\mathbf{B}$ as the product of 
user and item embedding matrices. i.e.,  $y_{ui}\approx\mathbf{E}_u\mathbf{E}_i^\top$ (denoted as element-wise), 
where $\mathbf{E}_u, \mathbf{E}_i\in\mathbb{R}^{1 \times d}$  are the representations of user $u$ and item $i$, respectively.


From another perspective, graph-based CF commonly takes user-item interactions $\mathcal{O^+}$ as the user-item bipartite graph $\mathcal{G}(\mathcal{V},\mathbf{A})$, where $\mathcal{V}=\mathcal{U} \cup \mathcal{I}$ and  
$\mathbf{A}$ denotes the  adjacency matrix  of $\mathcal{G}$: 
\begin{equation}
\mathbf{A}= \left[ {\begin{array}{*{20}{c}} 
\mathbf{0^{|\mathcal{U}|\times|\mathcal{U}|}}&\mathbf{B}^{|\mathcal{U}|\times|\mathcal{I}|}\\
{({\mathbf{B}^\top})^{|\mathcal{I}|\times|\mathcal{U}|}}&\mathbf{0^{|\mathcal{I}|\times|\mathcal{I}|}}
\end{array}} \right]
\end{equation}
where $\mathbf{0}$ denotes the null matrix. Therefore, implicit recommendations are naturally translated into link prediction in the user-item graph.
Inspired by the general architecture of GNN, graph-based CF methods mainly consist of Neighborhood Aggregation (NA) and Layer  Combination (LC), where NA is designed to aggregate the neighboring representations to update the central nodes of each layer and LC  combines the  presentations from incremental NA layers to obtain the final ones. Despite many NA functions have been proposed, most of them have the following symmetric forms:
 \begin{align}
  \mathbf{E}_u^{(l)} = f_{NA}(\mathbf{E}_u^{(l-1)}, \{\mathbf{E}_i^{(l-1)}|i\in\mathcal{N}_u\})~\label{equ:na}, \\
  \mathbf{E}_i^{(l)} = f_{NA}(\mathbf{E}_i^{(l-1)}, \{\mathbf{E}_u^{(l-1)}|u\in\mathcal{N}_i\}). ~\label{equ:na_}  
 \end{align}
where $\mathbf{E}_u^{(l)}$, $\mathbf{E}_i^{(l)} \in \mathbb{R}^{1\times d}$ denote the representations of user  $u$ and item  $i$ at $l$-th NA layer. $\mathbf{E}_u^{(0)}$ and $\mathbf{E}_i^{(0)}$ are the  initial node embeddings. $\mathcal{N}_u=\{i| (\mathbf{B})_{ui}=1, i \in \mathcal{I} \}$ and $\mathcal{N}_i=\{u| (\mathbf{B})_{ui}=1, u \in \mathcal{U} \}$ are the neighboring node set of  $u$ and $i$, respectively. $f_{NA}(\cdot)$ is the customized NA  function, e.g., element-wise mean~\cite{berg2017graph} or weighted sum~\cite{ying2018graph}. Besides, NGCF~\cite{wang2019neural} is proposed to add affinity propagation to the  NA layer. Further, He et al.~\cite{he2020lightgcn} found that feature transformation and non-linear activation are redundant in the NA function and the performance became better after removing them.
After propagating $L$ times NA layers, the LC function combines the node representations of each layer to obtain the final ones: 
\begin{align}
\mathbf{E} &= f_{LC}(\mathbf{E}^{(0)}, \mathbf{E}^{(1)}, \cdots, \mathbf{E}^{(L)})
\end{align}
where $f_{LC}(\cdot)$ denotes the LC function. For example, GCMC~\cite{berg2017graph} and PinSage~\cite{ying2018graph}  take the last NA layer's representations  as output. Both NGCF~\cite{wang2019neural} and LR-GCCF~\cite{chen2020revisiting} use concatenation, while LightGCN~\cite{he2020lightgcn} and SGL~\cite{wu2021self} adopt the average pooling. 

\begin{table*}[t]
\small
\centering
\caption{Recommendation
accuracy and novelty of  graph-based CF, where $\uparrow$ means higher is better and $\downarrow$ represents the opposite. }
\label{tab:GCN_SOTA}
\begin{threeparttable}
\begin{tabular}{c|cc|cc|cc|cc|cc|cc}
\toprule 
&\multicolumn{4}{c}{Amazon-Movie}  &\multicolumn{4}{|c}{Amazon-Book}&\multicolumn{4}{|c}{Yelp2018}\\
\cline{2-13}
&Recall$\uparrow$&NDCG$\uparrow$&Nov$\uparrow$&PRU$\downarrow$&Recall$\uparrow$&NDCG$\uparrow$&Nov$\uparrow$&PRU$\downarrow$&Recall$\uparrow$&NDCG$\uparrow$&Nov$\uparrow$&PRU$\downarrow$\\
\midrule 
MFBPR~\cite{rendle2009bpr}&0.0372&0.0241&0.5967&0.1120&0.0247&0.0193&0.6006&0.1581&0.0427&0.0347&0.5337&0.2129\\
\midrule 
GCMC~\cite{berg2017graph} 
&0.0482&0.0312&0.5844&0.1537&0.0284&0.0221&0.5990&0.1797&0.0435&0.0353&0.5698&0.0985\\
NGCF~\cite{wang2019neural} 
&0.0412&0.0268&0.5974&0.1104&0.0263&0.0203&0.6025&0.1485&0.0459&0.0371&0.5369&0.2074\\
LR-GCCF~\cite{chen2020revisiting} 
&0.0452&0.0291&0.5540&0.2133&0.0261&0.0202&0.5890&0.1840&0.0460&0.0377&0.5219&0.3114\\
LightGCN~\cite{he2020lightgcn} 
&0.0502&0.0323&0.5714&0.1854&0.0293&0.0225&0.5876&0.2118&0.0521&0.0426&0.5366&0.2485\\
SGL~\cite{wu2021self}
&0.0574&0.0368&0.5950&0.1652&0.0348&0.0268&0.5978&0.1648&0.0604&0.0492&0.5458&0.2311\\
\bottomrule 
\end{tabular}
     \end{threeparttable}
\end{table*}

\begin{table}[t]
\small
\centering
\caption{Novelty of LightGCN w.r.t propagation layer $L$.}
\label{tab:GCN_depth}
\begin{tabular}{c|c|c|c|c|c|c}
\toprule 

&\multicolumn{2}{c}{Amazon-Movie}  &\multicolumn{2}{|c}{Amazon-Book}&\multicolumn{2}{|c}{Yelp2018}\\
\cline{2-7}
&Nov$\uparrow$&PRU$\downarrow$&Nov$\uparrow$&PRU$\downarrow$&Nov$\uparrow$&PRU$\downarrow$\\
\midrule 
L=2 &0.5701&0.1756&0.5925&0.1839&0.5341&0.2464\\
L=4 &0.5714&0.1854&0.5900&0.2024&0.5366&0.2485\\
L=8&0.5684&0.2186&0.5783&0.2530&0.5175&0.3399\\
L=16&0.5425&0.3055&0.5534&0.3328&0.5128&0.3610\\
L=32&0.5119&0.3867&0.5193&0.4191&0.5071&0.3933\\
\bottomrule 
\end{tabular}
\vspace{-0.3cm}
\end{table}

During the prediction stage of  MF-based and graph-based CF, the preference  for unobserved interaction between user $u$ and item $i$ is given by $\hat{y}_{ui} = \mathbf{E}_u \mathbf{E}_i^\top$. Finally, the model parameters are optimized by the objective functions between the predicted scores and the ground truth. A common choice is to adopt BPR framework as the objective function \cite{rendle2009bpr}, which encourages higher scores for the items that users  have interacted with  than non-observed ones:
\begin{equation}\label{equ:bpr}
L_{BPR} = - \sum \limits_{(u,i,j)\in T} \ln\sigma(\hat{y}_{ui}-\hat{y}_{uj})+\lambda ||\mathbf{E}||_F^2.
\end{equation}
where $T = \{(u,i,j)| (i\in \mathcal{N}_u) \land (j\notin \mathcal{N}_u)\}$, $\sigma(\cdot)$ is the logistic sigmoid  and $\lambda$ controls the $L_2$ regularization coefficient. 


In summary, MF-based CF can be regarded as a special case of graph-based CF, where the depth of NA layer is 0. The main difference between them is that graph-based CF aggregates the high-order interactions in the embedding space, which is also known as the key for improving the accuracy. However, as the user-item interactions  are unbalanced with a power-law distribution, MF-based CF suffers the popularity bias since the popular items are more sufficiently optimized than  the novel ones. As for graph-based CF,  on the one hand,  NA layer can help users explore more diverse items. On the other hand, it can also enable the popular items more frequently reached by users. Therefore, whether graph-based CF can solve the popularity bias is an open question to be explored. 

\vspace{-0.1cm}
\subsection{Empirical Investigation on Accuracy and Novelty  of Graph-based CF}\label{sec:pop}
We first conduct experiments to investigate the two-fold performances w.r.t accuracy and novelty of the SOTA graph-based CF methods. 
In this work, we follow the commonly  accepted protocol of Top-K recommendation, where K=20\footnote{We also experiment with different Top-K values, and the overall trend is the same as Top-20. Due to the page limit, we only report K=20.}.  We use widely adopted Recall@K and NDCG@K to measure accuracy and use Nov@K~\cite{zhou2010solving} and PRU@K~\cite{zhu2021popularity}  (see Section~\ref{sec:data} for details) to measure novelty, respectively. Nov@K is defined as: 
\begin{equation}\label{equ:nov}
    Nov@K=\frac{1}{|\mathcal{U}|K}\sum\limits_{u\in \mathcal{U}}\sum\limits_{i\in I_u(K)}-\frac{1}{log_2|\mathcal{U}|} log_2(\frac{d_{i}}{|\mathcal{U}|}).
\end{equation}
where $I_u(K)$ denotes the Top-K ranked items for user $u$ during  testing  and $d_{i}$ denotes the number of observed interactions of item $i$ in the training set. 
Recently, researchers proposed to assess the correlation between the rank positions and popularity of top-ranked items recommended for users~\cite{zhu2021popularity}, i.e.,  PRU@K.  It is defined as:
\begin{equation}
    PRU@K=-\frac{1}{|\mathcal{U}|}\sum\limits_{u \in \mathcal{U}}SRC(\{d_{i}, rank_{ui}| i\in I_u(K)\}).
\end{equation}
where $SRC(\cdot,\cdot)$ calculates Spearman's rank correlation coefficient and $rank_{ui}$ returns the rank position of item $i$ for user $u$. 
We carefully fine-tine the hyper-parameters to ensure the best performance for each method~\footnote{Please note that the datasets, including Amazon-Book and Yelp2018, released by LightGCN DO NOT contain validation sets. In order to correctly implement early stopping  in the validation set,  we re-partition the training datasets so that the ratio of training set, validation set and test set is about 7:1:2.}. 
As shown in Table~\ref{tab:GCN_SOTA}, most graph-based recommendation methods~(except for NGCF) can achieve better prediction  accuracy, but they also have lower Nov@20 and higher PRU@20 than MFBPR. 
By definition, lower Nov@K indicates that  the top ranked items have larger popularity, while higher PRU@K  means that the more popular the items, the higher their ranking positions. As a whole,  the experiments show that most existing graph-based CF methods tend to recommend popular items to users, leading to a significant popularity bias. 
Moreover, many GNN-based methods are known to suffer accuracy degradation as the GNN layer deepens, possibly due to the over-smoothing problem~\cite{li2018deeper,chen2020measuring}. Therefore, we perform experiments to study the impact  of increasing NA layers on recommendation novelty. We report the results of LightGCN at different propagation layers $L$ in Table~\ref{tab:GCN_depth}. It can be found that as the depth of NA layer increases,  the performances of  Nov and PRU  deteriorate. It further indicates that LightGCN suffers from the more serious popularity bias as the NA layer goes deeper. Based on the above empirical evidences, we draw the following conclusions: 
\begin{itemize}
    \item  Existing graph-based CF methods prefer to recommend popular items  than niche ones, suffering the popularity bias. 
    
    \item As the depth of GNN layer increases, the popularity bias of graph-based CF  becomes more serious.
\end{itemize}
Although the emerging graph-based CF methods have great advantages in prediction accuracy, the low novelty caused by popularity bias should not be ignored. Except for the  over-smoothing problem, we empirically find that deep graph-based CF also suffers from the popularity bias, which is even more serious than shallow models.  This phenomenon is counter-intuitive, since the high-order propagation is good for users to explore unknown  interactions with items and brings more collaborative signals for optimization, but it also can exacerbate the imbalance between hot and long-tail items.  




\section{The Proposed Method}
Based on the investigations of Section~\ref{sec:pop}, 
we first analyse the causes of popularity bias in graph-based CF and then propose a simple yet effective method to increase recommendation novelty. 
\subsection{Theoretical Analyses}\label{sec:theorem}
Recall the following theorem for the analysis of over-smoothing issue in deep GNN~\cite{li2018deeper,liu2020towards,zhao2021bilateral}:
\begin{theorem}
Given a connected graph $\mathcal{G} (\mathcal{V}, \mathcal{E})$ with adjacency matrix $\mathbf{A}$. If  each node of  $\mathcal{G}$ is added self-loop, then
\begin{equation}
 \lim \limits_{l\to +\infty}(\tilde{{\mathbf D}}^{-r}\tilde{{\mathbf A}} {\tilde{\mathbf D}^{-(1-r)})^l_{i,j} =\frac{(d_i+1)^{1-r}(d_j+1)^r}{2|\mathcal{E}|+|\mathcal{V}|}}.
\end{equation}
where r $\in \mathbb{R} $, $\tilde{\mathbf{A}}=\mathbf{A}+\mathbf{I}$ , $\tilde{\mathbf{D}}=diag(d_1+1,...,d_{|\mathcal{V}|}+1)$ and $d_i=\sum_j \mathbf{A}_{ij}$.
\end{theorem}
We denote $\tilde{\mathbf{D}}^{-r}\tilde{\mathbf{A}}\tilde{\mathbf{D}}^{-(1-r)}$ as $r$-$AdjNorm$, where $r$ is the normalization coefficient. Due to  SOTA accuracy of LightGCN,  we use $r$-$AdjNorm$ as the NA function in Equations~(\ref{equ:na}) and~(\ref{equ:na_}). After propagating infinite times,  the node embedding matrix is updated to:
\begin{equation}\label{equ:gcn}
 \mathbf{H}^{(\infty)} =
(\tilde{\mathbf{D}}^{-r}\tilde{\mathbf{A}}\tilde{\mathbf{D}}^{-(1-r)})^\infty
\mathbf{H}^{(0)}.
\end{equation}
where $\mathbf{H}^{(l)}=[\mathbf{h}_1^{(l)}, \mathbf{h}_2^{(l)},..., \mathbf{h}_{|\mathcal{V}|}^{(l)}]^{\top}$ and $\mathbf{h}_1^{(l)}\in \mathbb{R}^{d \times 1}$.
Equation~(\ref{equ:gcn}) can be expressed in vector form:
\begin{equation}\label{equ:emb}
\begin{aligned}
\mathbf{h}_i^{(\infty)} = \frac{(d_i+1)^{1-r}}{2|\mathcal{E}|+|\mathcal{V}|}[(d_1+1)^r\mathbf{h}_1^{(0)}+(d_2+1)^r\mathbf{h}_2^{(0)}+\\\cdots+(d_{|\mathcal{V}|}+1)^r\mathbf{h}_{|\mathcal{V}|}^{(0)}].
\end{aligned}
\end{equation}
Researchers have demonstrated that inserting self-connections into the adjacent matrix is  equivalent to the LC function with a weighted sum~\cite{he2020lightgcn}.  During the inference phase, graph-based CF methods commonly use dot product to find the nearest items for the specific users. Given node $i$'s embedding as Equation~(\ref{equ:emb}), the dot product with node $j$  is:
\begin{equation}\label{equ:dot}
\begin{aligned}
(\mathbf{h}_i^{(\infty)})^{\top}\mathbf{h}_j^{(\infty)}=\frac{[(d_i+1)(d_j+1)]^{1-r}}{(2|\mathcal{E}|+|\mathcal{V}|)^2}[(d_1+1)^r\mathbf{h}_1^{(0)}+\\\cdots+(d_{|\mathcal{V}|}+1)^r\mathbf{h}_{|\mathcal{V}|}^{(0)}]^2.
\end{aligned}
\end{equation}
\begin{figure}[t] 
\centering 
\includegraphics[width=0.48\textwidth]{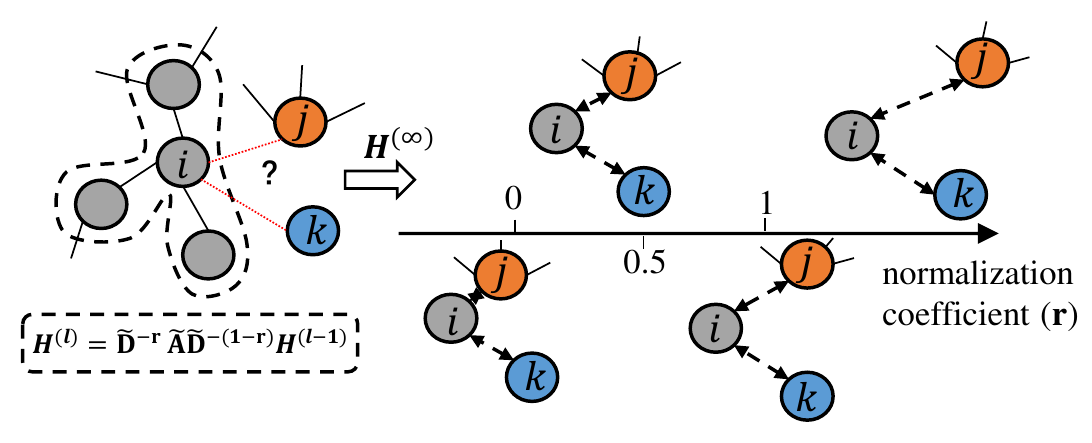} 
\caption{The trend of preference scores of the ego node $i$ with other nodes  as $r$ increases, where node $j$ has larger degree than node $k$. The length of the two-way arrow between  nodes represents the preference score and the shorter the length, the higher the score. It can be observed that ego node $i$ prefers node $j$ at first and then turns to node $k$ with the increase of $r$.  
} 
\label{pic:gcn_pop} 
\end{figure}
Since  $d_i$ is a positive integer,  Equation~(\ref{equ:dot}) has the following cases:
\begin{enumerate}
    \item\label{case:1} when $r<1$, if $d_j > d_k$ , then $(\mathbf{h}_i^{(\infty)})^{\top}\mathbf{h}_j^{(\infty)} >  (\mathbf{h}_i^{(\infty)})^{\top}\mathbf{h}_k^{(\infty)}$. That means node $i$ has larger dot product with node $j$ than that with node $k$ since the degree of node $j$ is larger than $k$. 
    \item when $r=1$, then $(\mathbf{h}_i^{(\infty)})^{\top}\mathbf{h}_j^{(\infty)} \equiv (\mathbf{h}_i^{(\infty)})^{\top}\mathbf{h}_k^{(\infty)}$. That is,  node $i$ has equal dot product with node $j$ and $k$, respectively, regardless of the degrees of node $j$ and $k$.
    \item when $r>1$, if $d_j > d_k$, then $(\mathbf{h}_i^{(\infty)})^{\top}\mathbf{h}_j^{(\infty)} < (\mathbf{h}_i^{(\infty)})^{\top}\mathbf{h}_k^{(\infty)}$. It implies that the smaller the node degree, the larger the dot product. That is the opposite of case~(\ref{case:1}).
\end{enumerate}
Based on the above analyses, the existing normalization forms of the adjacent matrix adopted in NA layer  can be classified as follows:
\begin{enumerate}
    \item\label{case:gn} As the most widely adopted symmetric normalization form, i.e., $\tilde{\mathbf{D}}^{-\frac{1}{2}}\tilde{\mathbf{A}}\tilde{\mathbf{D}}^{-\frac{1}{2}}$ ~\cite{wang2019neural,he2020lightgcn,chen2020revisiting,wu2021self}, 
    tends to  prioritize the nodes that have large degrees, probably leading to the bias towards popular items in the recommender systems.  Under this case, $r=0.5$ in Equation~(\ref{equ:gcn}) and it is denoted as $0.5$-$AdjNorm$.
    \item For right normalization $\tilde{\mathbf{A}}\tilde{\mathbf{D}}^{-1}$, i.e., $0$-$AdjNorm$ , it shows a similar trend with case~(\ref{case:gn}), but is more serious because it have a lower $r$ value. As a transition matrix for random walk in graph, 
    it is widely used for calculating the personalized PageRank~\cite{page1999pagerank, klicpera2018predict}.
    \item  $\tilde{\mathbf{D}}^{-1}\tilde{\mathbf{A}}$ , i.e., $1$-$AdjNorm$,  shows no bias towards node degree since $r=1$.  
    Wu et al.~\cite{wu2019neural} adopted the left normalization for social recommendation. This form has also been tested and found to be sometimes better than symmetric one~\cite{he2020lightgcn}.
    \item  $\tilde{\mathbf{D}}^{-r}\tilde{\mathbf{A}}\tilde{\mathbf{D}}^{-(1-r)}$ gives priority to recommend the low-degree nodes (a.k.a., long-tail nodes) when $r>1$.  A similar form is introduced \cite{chen2020scalable}, but we do not limit the range of $r$ and  further find out different properties for different intervals.
\end{enumerate}
To sum up, we can control the preference scores (measured by dot product) of the testing node pairs under graph-based CF by simply tuning the normalization coefficient $r$.   
Based on the theoretical analysis, we illustrate the preference scores of the ego node on candidate nodes with different  degrees in Figure~\ref{pic:gcn_pop}.
It can be found that ego node $i$ has a larger preference score for high-degree node $j$ when $r<1$. But the  score of ego node $i$ for low-degree node $k$ has smaller decay rates than that of high-degree one $j$ with the increase of $r$ value. When $r$ reaches the turning point, i.e., $r=1$, ego node $i$ has equal scores with node $j$ and $k$. Finally, the score for node $k$ becomes larger than $j$ when $r>1$, showing a preference for low-degree nodes. We will conduct experiments to confirm these findings in the next section. 
\subsection{$r$-$AdjNorm$  for  Graph-based CF backbones}\label{sec:backbone}
As shown above,  $r$-$AdjNorm$ can be easily plugged into the existing GNN-based CF backbones without any  additional computation. We take  LR-GCCF~\cite{chen2020revisiting} and LightGCN~\cite{he2020lightgcn} as two representative backbones, since both of them are  graph-based CF models that are specifically designed for the user-item bipartite graph structure, and have shown SOTA performances.  Besides,  both of them are simplifications for the GNN-based CF with different functions in the NA and LC layers, respectively. This property can help test the generality of our proposed plugin.

For LR-GCCF, it applies a symmetric normalized adjacent matrix with self-connections in the NA layer. Thus, the vector form of $r$-$AdjNorm$ plugin for LR-GCCF is:
 \begin{align}
  \mathbf{E}_u^{(l)} &= \frac{1}{|\mathcal{N}_{u}|}\mathbf{E}_u^{(l-1)} + \sum_{i\in\mathcal{N}_u}\frac{1}{{|\mathcal{N}_u|}^r{|\mathcal{N}_i|^{1-r}}}\mathbf{E}_i^{(l-1)}\label{equ:lrgccf_na}, \\
 \mathbf{E}_i^{(l)} &=  \frac{1}{|\mathcal{N}_{i}|}\mathbf{E}_i^{(l-1)} +  \sum_{u\in\mathcal{N}_i}\frac{1}{{|\mathcal{N}_i|}^r{|\mathcal{N}_u|}^{1-r}}\mathbf{E}_u^{(l-1)}.\label{equ:lrgccf_na_}  
 \end{align}
The value of $r$ controls the normalization strengths of the ego node and its neighborhood in the NA layer. We can obtain various NA functions by tuning $r$, e.g.,  the original propagation matrix of LR-GCCF is a special case of $r=0.5$. 
Combining Equations~(\ref{equ:lrgccf_na}) and~(\ref{equ:lrgccf_na_}),  the matrix form of the $r$-$AdjNorm$ plugin at $l$-th NA layer is:
\begin{equation}
\mathbf{E}^{(l)}  = {\tilde{\mathbf{D}}}^{-r}{\tilde{\mathbf{A}}}{\tilde{\mathbf{D}}}^{-(1-r)} \mathbf{E}^{(l-1)}, \mathrm{with}~ \tilde{\mathbf{A}}= \left[ {\begin{array}{*{20}{c}} 
\mathbf{I}&\mathbf{B}\\
{{\mathbf{B}^\top}}&\mathbf{I}
\end{array}} \right].
\end{equation}
where $\tilde{\mathbf{D}}$ is a diagonal degree matrix with $\tilde{D}_{ii}=\sum_j \tilde{\mathbf{A}}_{ij}$; $\mathbf{E}^{(l)} \in \mathbb{R}^{(|\mathcal{U}|+|\mathcal{I}|)\times d}$  is the embedding matrix for all users and items at $l$-th NA layer. After propagating NA layer $L$ times, high-order interactions between users and items are captured to represent the users' interests. Then, similar to the LC of LR-GCCF, the representations of each  layer are concatenated to obtain the final ones: 
\begin{align}
\mathbf{E} &= \mathbf{E}^{(0)} ||  \mathbf{E}^{(1)} || ...|| \mathbf{E}^{(L)}.
\end{align}
where $||$ denotes concatenation. 

For LightGCN, it  applies a symmetric normalized adjacent matrix without self-connections as the NA function. Therefore, the vector form of $r$-$AdjNorm$ plugin for LightGCN backbone is:
 \begin{align}
  \mathbf{E}_u^{(l)} &=  \sum_{i\in\mathcal{N}_u}\frac{1}{{|\mathcal{N}_u|}^r{|\mathcal{N}_i|^{1-r}}}\mathbf{E}_i^{(l-1)}, \\
 \mathbf{E}_i^{(l)} &=  \sum_{u\in\mathcal{N}_i}\frac{1}{{|\mathcal{N}_i|}^r{|\mathcal{N}_u|}^{1-r}}\mathbf{E}_u^{(l-1)}.  
 \end{align}
Likewise, the original propagation matrix of LightGCN is a special case of $r=0.5$. It should be noted that $r$-$AdjNorm$ of LightGCN doesn't include self-connections, which is different from that of LR-GCCF, for keeping consistent with the original backbone. 
We will empirically evaluate the simplification in the experimental part. The matrix form of $l$-th NA layer is:
\begin{equation}
  \mathbf{E}^{(l)}  = {\mathbf{D}}^{-r}{\mathbf{A}}{\mathbf{D}}^{-(1-r)} \mathbf{E}^{(l-1)}, \mathrm{with}~ \mathbf{A}= \left[ {\begin{array}{*{20}{c}} \mathbf{0}&\mathbf{B}\\
{{\mathbf{B}^\top}}&\mathbf{0}
\end{array}} \right].
\end{equation}
where $\mathbf{D}$ is a diagonal degree matrix with   $D_{ii}=\sum_j \mathbf{A}_{ij}$.  Then, similar to the LC function of LightGCN, the high-order representations of each NA layer are averaged to obtain the final ones for prediction:
\begin{align}
\mathbf{E} &= \frac{1}{L+1} (\mathbf{E}^{(0)}+  \mathbf{E}^{(1)} + ...+ \mathbf{E}^{(L)}).
\end{align}
Note that we keep the LC function unchanged for each backbone to highlight the influences of $r$-$AdjNorm$ in the NA layer.  Intuitively, the combination  coefficients in LC  should be carefully modeled as the depth of NA layer increases because deep GNN models are more prone to suffer over-smoothing and popularity bias than shallow GNN. In the paper,  we focus on studying the influence of NA layer  and leave the exploration of the mechanism for LC as future work.
\subsection{Model Training}
After obtaining the combined representations for all users and  items, we predict the preference of user $u$ for item $i$ as: $\hat{y}_{ui} = (\mathbf{E}_u)\mathbf{E}_{i}^{\top}$. Then we adopt the BPR framework  to optimize the model parameters. In the practical experiments,  the original BPR loss will lead to NAN errors, so the loss function is changed to the following safer form:
\begin{equation}
L_{BPR} = \frac{1}{|T|}\sum \limits_{(u,i,j)\in T}  softplus(-\hat{y}_{ui}+\hat{y}_{uj})+\lambda ||\mathbf{E}||_F^2.
\end{equation}
where $softplus(\cdot)$=$log(1+exp(\cdot))$ and $T$ is the same as Equation~(\ref{equ:bpr}).

\section{EXPERIMENTS}



\subsection{Experimental Settings}
\subsubsection{Datasets and Metrics}\label{sec:data}
In order to be consistent with previous researches~\cite{wang2019neural,he2020lightgcn,wu2021self}, we conduct experiments in three benchmark datasets, including Amazon-Movie~\cite{mcauley2015image}, Amazon-Book~\cite{mcauley2015image} and Yelp2018~\cite{wang2019neural}. For Amazon-Movie\footnote{\url{https://jmcauley.ucsd.edu/data/amazon}}, we filter out the users and items with less than 10 interactions.  For Amazon-Book and Yelp2018, we use the released data by LightGCN\footnote{\url{https://github.com/kuandeng/LightGCN}}. The statistics of the datasets are shown in Table~\ref{tab:datasets} and all the datasets are highly sparse. As mentioned in Section~\ref{sec:pop}, we redivide the datasets into training, validation and test set with a ratio of nearly  7:1:2.

In this work, we aim to investigate the two-fold performances of the graph-based CF methods. Therefore, we adopt four widely used metrics to measure the accuracy and novelty, respectively. To measure accuracy, we use Recall@K and NDCG@K where K=20. 
To measure novelty,  Zhou et al. \cite{zhou2010solving} used Surprisal/Novelty to measure the ability to recommend novel items. Surprisal/Novelty  is a widely used metric for recommender systems~\cite{vargas2011rank}. Since its range is  greater than 1, we normalize it as Equation~(\ref{equ:nov}). 
By definition, the less popular items are, the higher Nov will be. We also use the recently proposed PRU@K~\cite{zhu2021popularity} as a supplement. 
PRU@K measures the correlation between the items' rank positions and their popularity. The lower the PRU, the less bias toward popularity.

\begin{table}[t]
\small
\centering
\caption{Statistics of the datasets.}
\label{tab:datasets}
\begin{tabular}{c|c|c|c|c}
\toprule 
Datasets& \#Users & \#Items & \#Interactions& Sparsity\\
\midrule 
Amazon-Movie&40,928&51,509&1,163,413&0.0552\%\\
Amazon-Book&52,643&91,599&2,984,108&0.0619\%\\
Yelp2018&31,668&38,048&1,561,406&0.130\%\\
\bottomrule 
\end{tabular}
\end{table}

\subsubsection{Baselines}
To demonstrate the effectiveness of our proposed method, we compare it with the following  methods:  


\begin{table*}[t]
\small
\centering
\vspace{-0.1cm}
\caption{Overall performances w.r.t accuracy and novelty  of   competing methods under the backbone of LR-GCCF.}
\label{tab:LRGCCF}
\vspace{-0.1cm}
\begin{threeparttable}
\begin{tabular}{l|cc|cc|cc|cc|cc|cc}
\toprule 

&\multicolumn{4}{c}{Amazon-Movie}  &\multicolumn{4}{|c}{Amazon-Book}&\multicolumn{4}{|c}{Yelp2018}\\
\cline{2-13}
&Recall$\uparrow$&NDCG$\uparrow$&Nov$\uparrow$&PRU$\downarrow$&Recall$\uparrow$&NDCG$\uparrow$&Nov$\uparrow$&PRU$\downarrow$&Recall$\uparrow$&NDCG$\uparrow$&Nov$\uparrow$&PRU$\downarrow$\\
\midrule 
MFBPR&0.0372&0.0241&0.5967&0.1120&0.0247&0.0193&0.6006&0.1581&0.0427&0.0347&0.5337&0.2129\\
LR-GCCF&0.0452&0.0291&0.5540&0.2133&0.0261&0.0202&0.5890&0.1840&0.0460&0.0377&0.5219&0.3114\\
\midrule 
LR-GCCF+NS&0.0445&0.0289&0.5815&0.2153&0.0249&0.0196&0.6273&0.1771&0.0435&0.0362&0.5450&0.2880\\
LR-GCCF+PPNW&0.0427&0.0277&0.5961&0.1571&0.0256&0.0201&0.6021&0.1715&0.0440&0.0365&0.5505&0.2225\\
LR-GCCF+Reg&0.0390&0.0243&0.5694&0.0938&0.0234&0.0184&0.5976&0.1427&0.0437&0.0356&0.5378&0.2084\\
LR-GCCF+PC&0.0445&0.0284&0.5701&0.1157&0.0261&0.0203&0.5912&0.1455&0.0446&0.0356&0.5361&0.1851\\
LR-GCCF+MACR&0.0437&0.0280&0.5888&0.1923&0.0252&0.0194&0.6021&0.2487&0.0407&0.0325&0.5716&0.2064\\
LR-GCCF+DegDrop&0.0441&0.0282&0.5675&0.1764&0.0257&0.0200&0.5897&0.1849&0.0454&0.0377&0.5314&0.2673\\
\midrule 
LR-GCCF$_{1-AdjNorm}$&0.0463&0.0300&0.5838&0.1444& 0.0271&0.0210&0.6108&0.1545&0.0452&0.0370&0.5374&0.2286\\
LR-GCCF$_{0-AdjNorm}$&0.0390&0.0246&0.5334&0.2638&0.0242&0.0188&0.5551&0.2894&0.0449&0.0361&0.5123&0.3524\\
\midrule 
LR-GCCF\textbf{$_{r-AdjNorm}$}&0.0444&0.0287&0.6042&0.1163&0.0269&0.0205&0.6171&0.1460&0.0444&0.0359&0.5596&0.1478\\
\midrule 
\midrule 
\makecell[c]{\%Improv.\tnote{+}}&-0.22\%&1.06\%&5.98\%&-0.52\%&3.07\%&0.99\%&4.38\%&0.34\%&-0.45\%&0.84\%&3.98\%&20.15\%\\
\makecell[c]{$p$-$value$}&0.333&0.0608&1.39e-3\tnote{*}&0.430&0.0351\tnote{*}&0.211&4.92e-4\tnote{*}&0.487&0.334&0.102&1.47e-3\tnote{*}&1.79e-2\tnote{*}\\
\bottomrule 
\end{tabular}
\begin{tablenotes}
\footnotesize
\item[+]   
The improvements are calculated between LR-GCCF$_{r-AdjNorm}$ and LR-GCCF+PC, as LR-GCCF+PC reaches the best  accuracy and novelty trade-off of all baselines.
\item[*]  
It denotes that the corresponding improvement has passed the significant test at the significance level of 0.05.
\end{tablenotes}
\end{threeparttable}
\vspace{-0.2cm}
\end{table*}

\begin{itemize}
    \item Negative Sampling (NS) \cite{mikolov2013distributed}. The BPR loss as Equation~(\ref{equ:bpr}) uniformly samples negative user-item interactions, which will cause negative samples to be less popular than positive ones. Thus, we apply NS to the BPR loss.  To be specific, the possibility of negative item $j$ being sampled is $p(j)\propto d_j^\alpha$, where we tune  $\alpha$ in $[0, 0.25,0.5,0.75, 1]$. 
    \item PPNW \cite{lo2019matching}. PPNW considers personalized novelty weighting in the BPR loss. We use suggested values 
    of hyper-parameters by the authors and tune $\gamma$ in $[1,1.5,...,5] $ with a step size of 0.5, since $\gamma$  controls the strengths of  novelty.
    \item Reg \cite{zhu2021popularity}. This is a regularization method that 
    can effectively penalize high PRU value. 
    Thus, we tune the regularization coefficient $\gamma$ in $[1, 10, 10^2, ..., 10^5]$.
    \item PC \cite{zhu2021popularity}. This is a post-processing approach that directly modifies the predicted scores by compensating for item popularity. We 
    tune the trade-off $\alpha$ in $[10^{-4}, 10^{-3}, ..., 10^{-1},1]$.
    \item MACR \cite{wei2021model}. Different from the above methods, MACR is a recently proposed  method based on counterfactual reasoning  for eliminating the popularity bias. 
    We use the default values released by the authors' codes\footnote{\url{https://github.com/weitianxin/MACR}} 
    and tune $c$ in $[10,20,...,50]$. 
    \item DegDrop.  Inspired by DropEdge~\cite{rong2019dropedge},  we also propose a graph-based baseline, namely  DegDrop, which is more likely to aggregate the low-degree nodes in the NA layer. 
    To be specific, the possibility of an edge $(u, i)$ being dropped is proportional to $\alpha * d_i^{-1}$ and  $\alpha$ in tuned in $[0.1,0.2,..., 0.9]$.
\end{itemize}
In summary, the baselines cover a broad range of the methods for achieving accuracy and novelty trade-off, 
such as re-sampling, re-weighting, regularization, re-ranking, causality and so on. Note that many of the above baselines are  designed for  MF-based CF,  they are model-agnostic and applicable to graph-based CF.

\subsubsection{Hyper-parameter Settings}
For all the experimental methods, we adopt BPR loss and set the negative sampling rate to 1. Following the settings of LightGCN, we use xavier initializer to initialize the model parameters and apply the Adam optimizer  with a learning rate of 0.001. The embedding size is fixed as 64. To avoid over-fitting, $L_2$ normalization is searched in $\{10^{-5},10^{-4},..,10^{-1}\}$. Moreover, we adopt the same early stopping strategy as LightGCN and set the maximum epoch to 1,000 and the training process will be terminated if Recall@20 on the validation dataset does not increase for 5 evaluation times. 
That means our goal is to examine the performance of novelty when accuracy is at its best. For all hyper-parameters of the baselines, we use the values suggested by the corresponding papers with carefully fine-tuning on the new datasets. For $r$-$AdjNorm$, we retain the parameters of  backbones and tune $r$ in  $[0.5, 0.55, ..., 1.5]$ with a 0.05 step-size. 

\begin{table*}[htbp]
\small
\centering
\caption{Overall performances w.r.t accuracy and novelty  of  competing methods under the backbone of LightGCN.}
\label{tab:LightGCN}
\begin{threeparttable}
\begin{tabular}{l|cc|cc|cc|cc|cc|cc}
\toprule 

&\multicolumn{4}{c}{Amazon-Movie}  &\multicolumn{4}{|c}{Amazon-Book}&\multicolumn{4}{|c}{Yelp2018}\\
\cline{2-13}
&Recall$\uparrow$&NDCG$\uparrow$&Nov$\uparrow$&PRU$\downarrow$&Recall$\uparrow$&NDCG$\uparrow$&Nov$\uparrow$&PRU$\downarrow$&Recall$\uparrow$&NDCG$\uparrow$&Nov$\uparrow$&PRU$\downarrow$\\
\midrule 
MFBPR&0.0372&0.0241&0.5967&0.1120&0.0247&0.0193&0.6006&0.1581&0.0427&0.0347&0.5337&0.2129\\
LightGCN&0.0502&0.0323&0.5714&0.1854&0.0293&0.0225&0.5876&0.2118&0.0521&0.0426&0.5366&0.2485\\
\midrule 
LightGCN+NS&0.0496&0.0323&0.6067&0.1755&0.0283&0.0219&0.6093&0.1705&0.0499&0.0404&0.5593&0.2088\\
LightGCN+PPNW&0.0493&0.0320&0.5946&0.1540&0.0284&0.0219&0.6063&0.1641&0.0503&0.0410&0.5492&0.2388\\
LightGCN+Reg&0.0465&0.0273&0.5490&0.1344&0.0276&0.0210&0.5823&0.1837&0.0488&0.0392&0.5324&0.1899\\
LightGCN+PC&0.0507&0.0324&0.5765&0.1404&0.0289&0.0221&0.5951&0.1491&0.0515&0.0414&0.5362&0.2314\\
LightGCN+MACR&0.0503&0.0318&0.5499&0.2877&0.0232&0.0183&0.4940&0.4151&0.0502&0.0408&0.5373&0.2701\\
LightGCN+DegDrop&0.0492&0.0320&0.5737&0.1734&0.0291&0.0224&0.5908&0.1966&0.0516&0.0424&0.5406&0.2273\\
\midrule 
LightGCN$_{1-AdjNorm}$&0.0512&0.0330&0.5910&0.1423&0.0299&0.0230&0.6013&0.1741&0.0516&0.0419&0.5438&0.2313\\
LightGCN$_{0-AdjNorm}$&0.0421&0.0264&0.5368&0.2694&0.0256&0.0196&0.5557&0.3216&0.0517&0.0416&0.5103&0.3972\\
\midrule 
LightGCN$_{r-AdjNorm}$&0.0504&0.0327&0.5959&0.1349&0.0296&0.0228&0.6136&0.1529&0.0514&0.0416&0.5559&0.1876\\
\midrule 
\midrule 
\makecell[c]{\%Improv.\tnote{+}}&-0.59\%&0.93\%&3.37\%&3.92\%&2.42\%&3.17\%&3.11\%&-2.55\%&-0.19\%&0.48\%&3.67\%&18.93\%\\
\makecell[c]{$p$-$value$}&0.289&0.0932&1.50e-5\tnote{*}&0.0129\tnote{*}&6.78e-3\tnote{*}& 4.29e-4\tnote{*}&1.09e-5\tnote{*}&0.0190\tnote{*}&0.272&0.172&1.09e-4\tnote{*}&7.50e-3\tnote{*}\\
\bottomrule 
\end{tabular}
\begin{tablenotes}
\footnotesize
\item[+]   The improvements are calculated between LightGCN$_{r-AdjNorm}$ and LightGCN+PC.
\end{tablenotes}
\end{threeparttable}
\vspace{-0.1cm}
\end{table*}

\subsection{Overall Performance Comparison}\label{sec:overall}
We conduct detailed experiments for comparison with benchmark methods under the backbones of LR-GCCF and LightGCN, which are denoted as LR-GCCF$_{r-AdjNorm}$ and LightGCN$_{r-AdjNorm}$ respectively. 
Since most existing baselines are seeking for a trade-off performance between accuracy and novelty, we fine-tune the hyper-parameters of each competing method to reach the novelty level of MFBPR and then observe the decline in accuracy. We report the results in Table~\ref{tab:LRGCCF} and Table~\ref{tab:LightGCN}, where all the results are the average values of 5 repeated runs. 
The main observations are  as follows: 
\begin{itemize}
    \item Most benchmark methods, including re-sampling (i.e. NS), re-weighting (i.e. PPNW), regularization (i.e. Reg), re-ranking (i.e. PC) and DegDrop, can improve Nov@20 or reduce  PRU@20  compared to the original backbones, but at the expense of sacrificing  accuracy. In the experiments, we find that all of these methods will lead to a decrease in  accuracy when improving novelty, showing a trade-off between them. We also find  that while Nov@20 and PRU@20 are correlated, it doesn't mean that both indicators can be improved simultaneously. 
    For example,  LightGCN+Reg can significantly improve PRU@20, but it results in a decrease in Nov@20 referring to Table~\ref{tab:LightGCN}. 
    While NS and PPNW perform better  than Reg on Nov@20 because they directly penalize item popularity  in the loss function, while Reg aims to penalize the correlation between the predicted scores of positive user-item pairs and item popularity. 
    In addition to the conventional data-driven methods, we also conduct experiments with the counterfactual reasoning framework (i.e., MACR). 
    We find that MACR doesn't perform well in our evaluation protocol after  fine-tuning the hyper-parameters. The reason may be that our test set is not  debiased data, while MACR is  evaluated to be effective on the debiased test set. As for DegDrop, it can  achieve a trade-off performance, 
    but we also find that it is not flexible enough to adjust the two-fold performances.
    That is, adjusting the dropout ratio $\alpha$ doesn't observe a significant change in novelty. Overall, PC performs best among the baselines and can  maintain  accuracy while improving novelty. 
    
    \item  Moreover, we  find that 1-$AdjNorm$  shows superior performances w.r.t accuracy and novelty than 0-$AdjNorm$ in both LR-GCCF and LightGCN. Especially, 1-$AdjNorm$ always has  higher Nov@20 and lower PRU@20 than  0-$AdjNorm$ in all datasets.  We also note that the testing novelty of the  backbones (0.5-$AdjNorm$) lies between the above two plugins. The results confirm the analysis in Section~\ref{sec:theorem} that $r$ controls the preference for nodes with different degrees, albeit the depth of NA layer doesn't reach infinity in practice.
    \item  Further, we fine-tune the normalization coefficient $r$ to make the original backbones  reach the same level of novelty as MFBPR. From Table~\ref{tab:LRGCCF} and  Table~\ref{tab:LightGCN}, $r$-$AdjNorm$  can significantly improve novelty without sacrificing accuracy compared to the original backbones. Meanwhile, it outperforms  the above  baselines in terms of a good accuracy and novelty trade-off. The improvement and  $p$-$value$ 
   are calculated with  the strongest baseline, i.e., PC. It shows that $r$-$AdjNorm$'s improvements in novelty are statistically significant, while in most cases there is no significant difference in their accuracy. We also find that the improvements of  $r$-$AdjNorm$ for LightGCN are  more outstanding than that of LR-GCCF, manifesting that self-connection is not a prerequisite.  
   \item Last but not least, the performances of different backbones w.r.t LR-GCCF and LightGCN  verify the generality of $r$-$AdjNorm$.  Note that it doesn't require any additional computation, only the normalized form of the adjacency matrix needs to be adjusted during the pre-processing. Thus, it provides an efficient and effective way for graph-based CF, especially one using degree normalization, to explore a trade-off between accuracy and novelty.

\end{itemize}

\subsection{Study of Hyper-parameters}
In the section, we investigate the impacts of different  hyper-parameters, including normalization coefficient $r$ and the depth of propagation layer $L$. We choose LightGCN as the backbone because of its better recommendation accuracy performance compared to LR-GCCF.

\begin{table*}[t]
\small
\centering
\vspace{-0.3cm}
\caption{Comparisons of trade-off performances between SGL and SGL$_{r-AdjNorm}$ under different hyper-parameters. }
\vspace{-0.3cm}
\label{tab:backbones}
\begin{threeparttable}
\begin{tabular}{c|cccc|cccc|cccc}
\toprule 
&\multicolumn{4}{c|}{Amazon-Movie} &\multicolumn{4}{c|}{Amazon-Book}  &\multicolumn{4}{c}{Yelp2018}\\
\cline{2-13}
&Recall$\uparrow$&NDCG$\uparrow$&Nov$\uparrow$&PRU$\downarrow$&Recall$\uparrow$&NDCG$\uparrow$&Nov$\uparrow$&PRU$\downarrow$&Recall$\uparrow$&NDCG$\uparrow$&Nov$\uparrow$&PRU$\downarrow$\\
\midrule 
SGL~(acc. opt.)&0.0574&0.0368&0.5950&0.1652&0.0348&0.0268&0.5978&0.1648&0.0604&0.0492&0.5458&0.2311\\
SGL~(nov. opt.)&0.0538&0.0358&0.6004&0.1308&0.0336&0.0263&0.6001&0.1444&0.0572&0.0475&0.5625&0.1698\\
\midrule
SGL$_{1-AdjNorm}$&0.0543&0.0357&0.6975&0.0695&0.0340&0.0262&0.6861&0.0601&0.0557&0.0454&0.5910&0.1473\\
SGL$_{0.75-AdjNorm}$&0.0572&0.0372&0.6300&0.1265&0.0348&0.0268&0.6361&0.1232&0.0594&0.0484&0.5706&0.1714\\
SGL$_{0.25-AdjNorm}$&0.0536&0.0341&0.5495&0.2586&0.0320&0.0247&0.5588&0.2355&0.0593&0.0485&0.5233&0.2986\\
SGL$_{0-AdjNorm}$&0.0471&0.0294&0.5057&0.3340&0.0256&0.0203&0.5063&0.3554&0.0557&0.0453&0.4820&0.3906\\
\bottomrule 
\end{tabular}
\end{threeparttable}
\vspace{-0.3cm}
\end{table*}

\begin{figure}[htbp] 
\centering 
\includegraphics[width=0.48\textwidth]{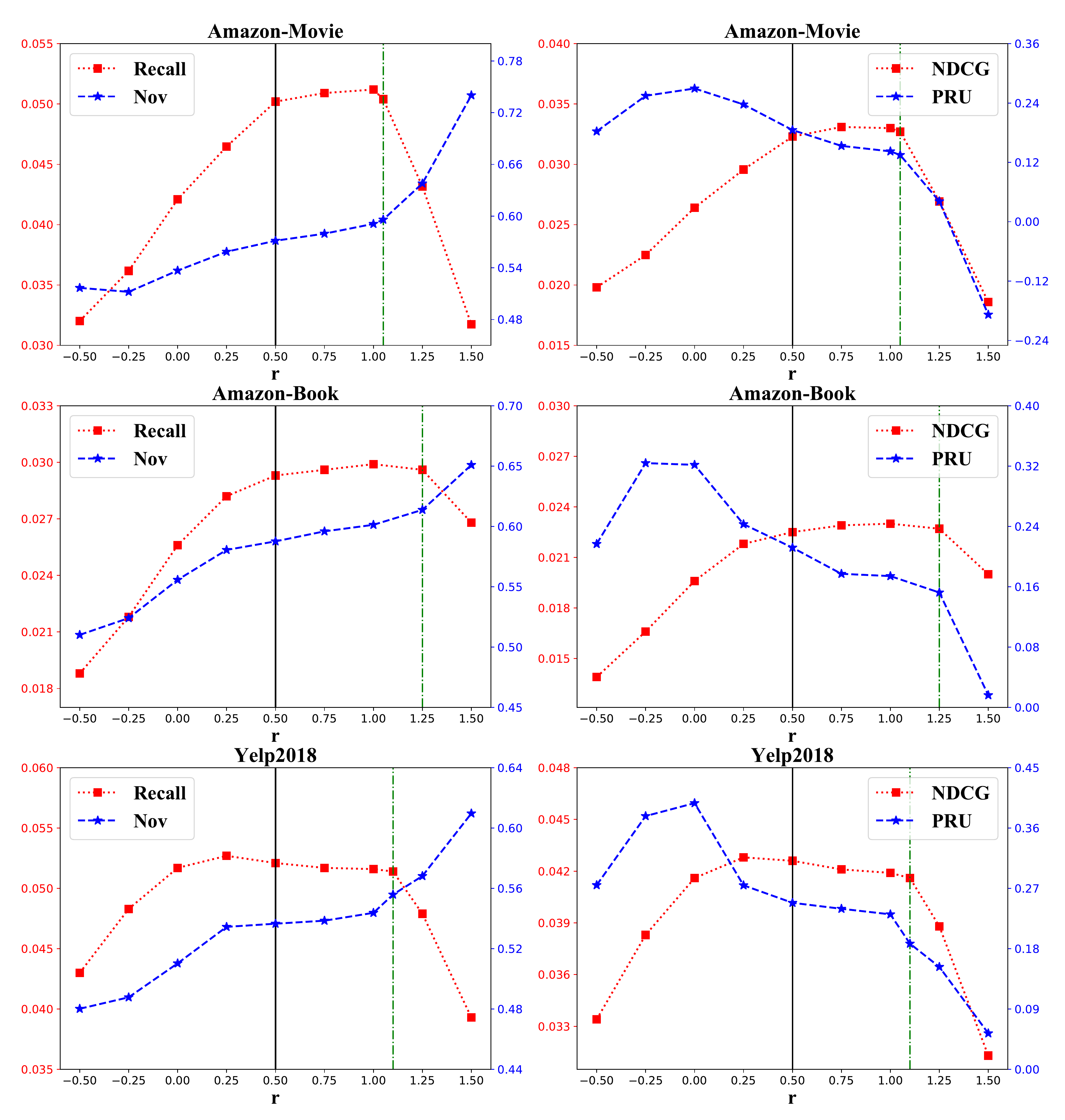} 
\vspace{-0.3cm}
\caption{Model performances w.r.t   normalization coefficient $r$ on three datasets. The black and  green vertical  lines represent the  original backbone and the fine-tuned one in Table~\ref{tab:LightGCN}. 
} 
\label{pic:r_gn} 
\vspace{-0.2cm}
\end{figure}

\begin{figure}[t] 
\centering 
\includegraphics[width=0.48\textwidth]{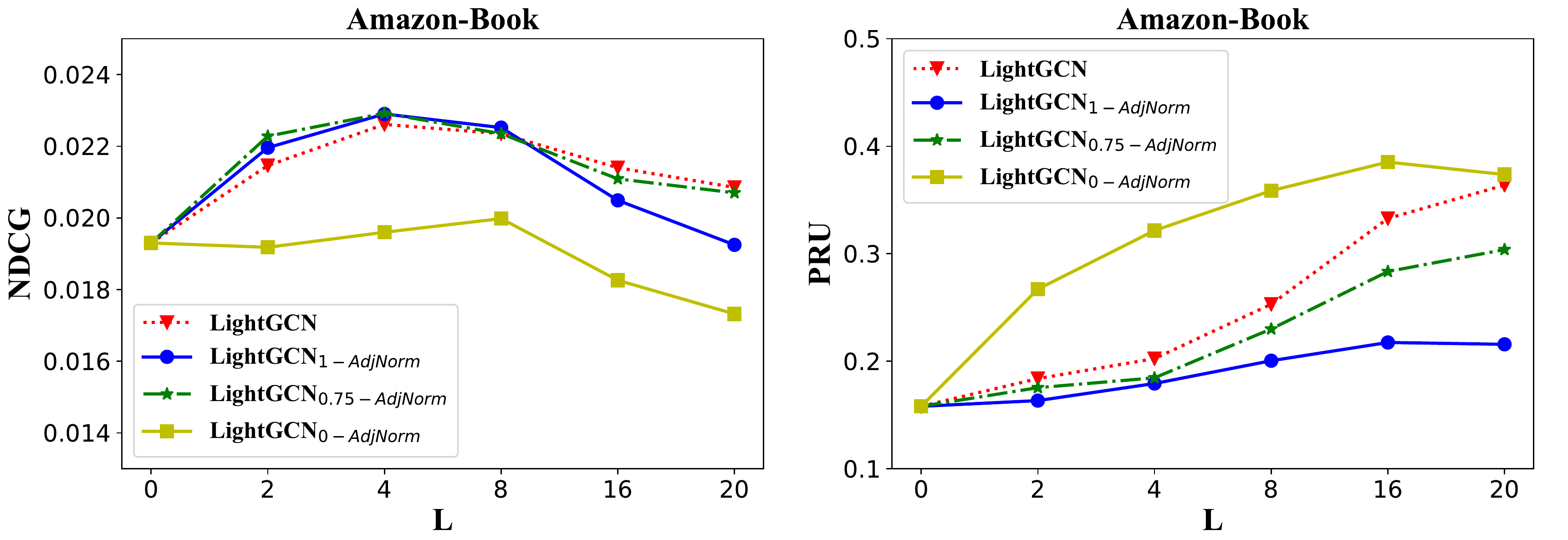}
\vspace{-0.2cm}
\caption{The accuracy and novelty performances of LightGCN$_{r-AdjNorm}$ w.r.t varying propagation layer $L$. 
}
\label{pic:pl} 
\vspace{-0.3cm}
\end{figure}

\emph{Effect of Normalization Coefficient $r$.}
As analyzed in Section~\ref{sec:theorem}, $r$ plays an important role in controlling the bias towards node degree. Therefore, we adjust $r$ to study the influences on the accuracy and novelty  and report the results in Figure~\ref{pic:r_gn}.  The  performances of original backbone are emphasized by  black vertical lines. We can find that both Recall and NDCG rise first then fall as  $r$ increases from -0.5 to 1.5 with step 0.25. On the contrary, Nov and PRU show a continuous  upward and downward trend, respectively, except for very few data points.  As for LightGCN, the two-fold performances w.r.t accuracy and novelty both are in the middle. When $r>1$, both Nov and PRU change rapidly, indicating a significant improvement in the novelty.
The phenomenon is consistent with the analysis in Section~\ref{sec:theorem}  that the preference of deep GNN-based CF for low-degree  nodes is enhanced as the normalization coefficient $r$ increases, which is reflected in an increase of Nov and a decrease in PRU. The experiment demonstrates that the proposed $r$-$AdjNorm$ plugin  can fine tuning  the accuracy and novelty of LightGCN in an efficient and effective way. 
In general, we can change the novelty performance  by setting different $r$ values. However, too large or too small values will cause  the model to favor popular or long-tail items too much and have negative effects on the model accuracy.
Hence, adjusting $r$ in a proper range can achieve a win-win situation where both accuracy and novelty are improved. According to  empirical results, we suggest to tune $r$ in the range of 0.5$\sim$1.25 carefully to achieve a good accuracy and novelty trade-off.

\emph{Effect of Propagation Layer $L$.}
In Section~\ref{sec:pop}, we have observed that  novelty metrics of LightGCN drop as the number of propagation layer increases. In this section, we  conduct experiments with different variants of $r$-$AdjNorm$ to  study the effects of propagation layer $L$ on both accuracy and novelty.  
As a special case, $L=0$ denotes the performances of MFBPR.
As can be seen from the Figure~\ref{pic:pl}, the NDCG of all variants of $r$-$AdjNorm$ 
first increases and then decreases with the increase of $L$. But $1$-$AdjNorm$ and 0-$AdjNorm$ have poor performances in NDCG when $L>8$ compared to LightGCN. At the same time, PRU shows an  upward trend (Please note that the lower the PRU, the better the novelty). To be specific, 1-$AdjNorm$  performs  best in terms of PRU, followed by 0.75-$AdjNorm$ and LightGCN (i.e, 0.5-$AdjNorm$), and 0-$AdjNorm$ performs the worst.  
In addition, the PRU value of $1$-$AdjNorm$ is  closest to MFBPR and  increases at a slower rate as L increases. 
Among them, 0.75-$AdjNorm$ can significantly improve the novelty of the LightGCN backbone without sacrificing  accuracy. We omit the Recall and Nov due to space limitation, while they show the same results.
As a whole, we can achieve a good accuracy and novelty trade-off of graph-based CF simply by applying $r$-$AdjNorm$.



\vspace{-2mm}
\subsection{Performance on  Self-supervised Graph Learning~(SGL) Backbone}

In previous experiments, we have verified the effectiveness of $r$-$AdjNorm$ based on the backbones of LR-GCCF and LightGCN, respectively. In this part, we study the effect of our proposed plugin for a SOTA Self-supervised Graph Learning~(SGL) model~\cite{wu2021self}. SGL is composed of two losses with a balance parameter: a classical supervised BPR loss from skewed user-item interaction, and an auxiliary self-supervised loss by randomly enforcing user~(item) representation learning via self-discrimination. Specifically, the self-supervised loss could alleviate the popularity bias as it randomly selects nodes in the user-item graph, and can make long-tail recommendation if we put more weights on the self-supervised loss. To investigate that whether our proposed $r$-$AdjNorm$ works on SGL backbone, we first fine-tune the balance parameter of SGL to achieve the optimal accuracy and optimal novelty, denoted as SGL (acc. opt.) and SGL (nov. opt.), respectively.  Then, we put our proposed plugin into SGL and adjust the values of normalization coefficient  $r$ to observe the performances. As is shown in Table~\ref{tab:backbones},
SGL shows better accuracy and novelty compared to other graph-based backbones, which is due to the enhancement of the self-supervised loss.
The accuracy of SGL degrades when its novelty is tuned to be optimal. Second,  SGL$_{0.75-AdjNorm}$ shows better accuracy and  novelty at the same time  compared to SGL~(nov. opt.), indicating that our proposed method can further achieve a good accuracy-novelty trade-off based on SGL and is more effective than simply tuning the balance parameter of SGL. Last but not least, the trends between $r$ and  novelty performances on the  SGL backbone are the same as theoretical analysis.  
This study shows  that the proposed $r$-$AdjNorm$ can be also flexibly integrated into SGL backbone.

\section{CONCLUSIONS}
In this work, we studied the accuracy and novelty  performances  of graph-based CF methods. We empirically found that most existing graph-based CF methods tend to exacerbate the popularity bias.
In particular, we theoretically analyzed the cause for this phenomenon  and  proposed an effective method by adjusting the normalization strengths in the NA process adopted by the current graph-based CF models. 
We conducted extensive experiments on three benchmark datasets to demonstrate the effectiveness of our proposed method regarding novelty improvement without sacrificing accuracy under various graph-based CF backbones. 

\section*{Acknowledgments}
This work was supported in part by grants from the National Key R\&D Program of China (Grant No. JZ2021ZD0111802), and the Young Elite Scientists Sponsorship Program by CAST and ISZS.

	{
		\bibliographystyle{ACM-Reference-Format}
		\balance
		\bibliography{acmart}
	}

\end{document}